\documentclass[fleqn,usenatbib]{mnras}

\usepackage{newtxtext,newtxmath}
\usepackage{deluxetable}
\newcommand\SPSB[2]{^{#1}_{#2}}
\usepackage[T1]{fontenc}
\usepackage{ae,aecompl}
\usepackage{pdflscape}

\usepackage{graphicx}	% Including figure files
\usepackage{amsmath}	% Advanced maths commands
\usepackage{amssymb}	% Extra maths symbols

\title[Biases in inferring dark matter profiles]{Biases in inferring dark matter profiles from dynamical and lensing measurements}

\author[S. Scibelli et al.]{
Samantha Scibelli,$^{1,2}$\thanks{E-mail: sscibelli@email.arizona.edu}
Rosalba Perna,$^{2,3}$
Charles Keeton$^{4}$
\\
$^{1}$Steward Observatory, University of Arizona,
Tucson, AZ 85721\\
$^{2}$Department of Physics and Astronomy,
Stony Brook University, 
Stony Brook, NY 11790\\
$^{3}$Center for Computational Astrophysics, Flatiron Institute, New York, NY 10010, USA\\
$^{4}$Department of Physics and Astronomy, 
Rutgers University, 
Piscataway, NJ 08854\\
}

\date{Accepted XXX. Received YYY; in original form ZZZ}

\pubyear{2018}

\begin{document}
\label{firstpage}
\pagerange{\pageref{firstpage}--\pageref{lastpage}}
\maketitle

\begin{abstract}
The degeneracy between disc and halo contributions in spiral galaxy rotation curves makes it difficult to obtain a full understanding of the distribution of baryons and dark matter in disc galaxies like our own Milky Way. Using mock data, we study how constraints on dark matter profiles obtained from kinematics, strong lensing, or a combination of the two are affected by assumptions about the halo model. We compare four different models: spherical isothermal and Navarro-Frenk-White halos, along with spherical and elliptical Burkert halos. For both kinematics and lensing we find examples where different models fit the data well but give enclosed masses that are inconsistent with the true (i.e., input) values. This is especially notable when the input and fit models differ in having cored or cuspy profiles (such as fitting an NFW model when the underlying dark matter distribution follows a different profile). We find that mass biases are more pronounced with lensing than with kinematics, and using both methods can help reduce the bias and provide stronger constraints on the dark matter distributions.
\end{abstract}


%%%%%%%%%%%%%%%%%%%%%%%%%%%%%%%%%%%%%%%%%%%%%%%%%%

%%%%%%%%%%%%%%%%% BODY OF PAPER %%%%%%%%%%%%%%%%%%

\section{Introduction} \label{sec:intro}

Some of the earliest evidence for dark matter (DM) came from spiral galaxies, whose rotation curves reveal the gravitational influence of unseen matter \citep[e.g.,][]{1978ApJ...225L.107R,1981AJ.....86.1825B}. As the Cold Dark Matter (CDM) paradigm emerged, $N$-body simulations showed that pure dark matter halos in equilibrium have spherically-averaged density profiles that are nearly universal and `cuspy' (i.e., the profile rises steeply toward the center; \citealt{1996ApJ...462..563N,2010MNRAS.402...21N,2016JCAP...09..042N}). By contrast, detailed studies of spiral galaxy rotation curves seem to favor `cored' profile with a shallow or even flat dark matter density profile at small radii \citep[e.g.,][]{2007MNRAS.378...41S,2011AJ....142...24O,2017MNRAS.465.4703K}. Baryonic feedback appears to be important in resolving this cusp/core problem: bursts of star formation and the associated feedback can generate repeated fluctuations in the central potential well, changing the density profile and turning cusps into cores \citep[e.g.,][]{2010ApJ...716..198B,2010MNRAS.405.2161D,2013MNRAS.429.3316B,2013MNRAS.429.3068T,2014MNRAS.437..415D,2015MNRAS.454.2981C,2017ApJ...850...97B}. Careful measurements of the inner profiles of dark matter halos could therefore probe the baryonic processes that occur during galaxy formation and evolution \citep[e.g.,][]{2012ApJ...750...10S}.

If the DM profile were known, we could use rotation curves to determine the mass-to-light ratio of the baryonic component and hence constrain the low-mass end of the stellar initial mass function (IMF). However, since the profile is not known, current analyses cannot fully disentangle the baryonic and DM contributions to the rotation curve. Some assumptions yield models with a relatively low-mass disc and a dense halo, while others lead to a more massive disc and less massive halo \citep[e.g.,][]{1986RSPTA.320..447V,2001MNRAS.325.1017V,2005ApJ...619..218D}.

One way to ameliorate this `disc-halo degeneracy' is to combine rotation curves with strong gravitational lensing, whenever both data are available for the same system \citep[e.g,][]{2000ApJ...533..194M,2002MNRAS.334..621T}. Since the disc and halo generally have different projected ellipticities, and thus different effects on lensing observables \citep{1998ApJ...495..157K}, lensing offers a complementary way to separate the stellar and DM components and break degeneracies that arise with kinematics alone \citep[e.g.][]{2011MNRAS.417.1621D,2011MNRAS.415.2215B,2018MNRAS.475.2403L}. Observational studies of spiral galaxy lenses have already been used to make inferences about disc masses and stellar IMFs \citep[e.g.,][]{2010MNRAS.401.1540T}.

When inferring physical parameters from models, the crucial question is whether finding a statistically good fit means the model is `correct' in the sense that the parameter values correspond to the underlying dark matter distribution. This is an issue that can be investigated only if the actual DM distribution is known independently. We create such a condition by generating mock data using four of the most commonly used DM profiles (Isothermal Sphere, Navarro-Frank-White, spherical Burkert and elliptical Burkert). The characteristics of the data (i.e., number of data points, measurement uncertainties) are chosen to be consistent with those typical of current surveys. Taking an agnostic view of what the actual underlying DM profile is, we fit all four mock datasets with all four models and examine how masses recovered from the models compare with the input masses. In this way we quantify the statistical uncertainties from typical kinematic and lensing investigations, and we examine whether adopting incorrect assumption about the dark matter halo model leads to biases in mass measurements. We consider kinematics and lensing separately and together in order to evaluate each method independently and ascertain whether a joint analysis yields improved results.

This paper is organized as follows. We summarize our kinematic methods in Sec.~\ref{sec:kinematics} and our lensing methods in Sec.~\ref{sec:lensing}. We describe the statistical analysis of fits in Sec.~\ref{sec:fitting}. We then present the fit results in Sec.~\ref{sec:results}, first for kinematics alone, then for lensing along, and finally for both together. Finally, we summarize and conclude with a discussion in Sec.~\ref{sec:conclusions}.

%%%%%%%%%%%%%%%%%%%%%%%%%%%%%%%%%%%%%%%%%%%%%%%%%%
\section{Rotation Curve Modeling} \label{sec:kinematics}
%%%%%%%%%%%%%%%%%%%%%%%%%%%%%%%%%%%%%%%%%%%%%%%%%%

For the kinematic analysis, we assume that stars and gas follow circular orbits and model the rotation curve, $v_c(R)$, with two contributions:
\begin{equation}
    v_c^2 = v_d^2 + v_h^2 ,
\end{equation}
where $v_d$ denotes the contribution from the disc, for which we use an exponential disc profile (Sec.~\ref{sec:ExpProf}), while $v_h$ denotes the contribution from the dark matter halo, for which we use different models as described below. We neglect any bulge component because bulges are typically well-constrained and display only mild correlations with both the disc and the halo \citep[e.g.,][]{2010MNRAS.401.1540T}.  Since it is the  disc-halo degeneracy that is the most prominent in observations, we focus our study on this. However, it will be interesting in future studies to consider including a bulge.

\begin{figure}
\begin{center}$
\begin{array}{c}
\includegraphics[width=80mm]
{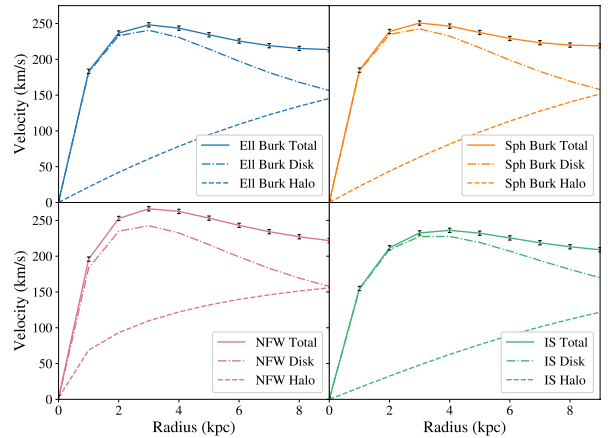}
\end{array}$
\end{center}
\caption{Rotation curves out to a distance of 9~kpc produced from mock data created from the elliptical Burkert profile in blue (top left), spherical Burkert profile in orange (top right), NFW in purple (bottom left), and IS halo in green (bottom right). In black we have plotted the total rotation curve error bars at each point along the curves, corresponding to 3~km~s$^{-1}$. 
}
\label{rotcurve1}
\end{figure}

The disc and halo parameters used to create the mock data are reported in Table \ref{mock}. For each combination of disc and halo, we produce mock rotation curve data out to 9~kpc, as shown in Figure \ref{rotcurve1}. This choice is motivated by the typical  data in studies that couple lensing and kinematics \citep[e.g.,][]{2010MNRAS.405.2161D, 2012ApJ...750...10S}. Instead of trying to compare each disc and halo parameter from model to model, which can be challenging when different models have different parametrisations, we compute the enclosed mass of the disc and halo in order to make meaningful comparisons. The enclosed mass values for each of the four mock data cases are presented below in Table \ref{mock_encl_mass}. 

\begin{table}
\centering
\caption{\label{mock}
Rotation curve parameters used to create the mock rotation data. Values loosely based on \citet{2004MNRAS.351..903G} but chosen to create galaxies massive enough to serve as strong lenses.
}
\centering
\begin{tabular}{lcccc}
\hline
%Models & disc params. &   &  halo params. &  \\
  & $r_d$  & $M$ & $r_c$ & $\rho_c$ \\
  &  kpc &  $10^{10}\ M_\odot$ & kpc & $10^7$ $M_\odot$/kpc$^3$ \\
\hline\hline
Ell. Burkert & 1.3 & 4.5 & 13.0 & 3.0 \\
Sph. Burkert & 1.3 & 4.5 & 13.0 & 3.0 \\
\hline
& $r_d$ & $M$  & $M_{\rm vir}$ & $\rho_s$ \\
& kpc & $10^{10}\ M_\odot$ & $10^{11}\ M_\odot$ & $10^7$ $M_\odot$/kpc$^3$ \\
\hline
NFW & 1.3 & 4.5 & 8 &  1.3 \\ 
 %NFW & 3.65e21   & 4.9e43 & 41 &  7.75e-31 \\  41 is real... unrealistic so we changed it
\hline 
 & $r_d$ (kpc) & $M$ & $r_c$ & $\rho_0$  \\
  & kpc & $10^{10}\ M_\odot$ & kpc & $10^7$ $M_\odot$/kpc$^3$\\
 \hline
  IS & 1.6 & 5.0 & 9.7 & 1.5\\
\hline
\end{tabular}
\end{table}

\begin{table}
\centering
\caption{\label{mock_encl_mass}
Total enclosed mass based on the mock data for the four models.
}
\begin{tabular}{lcccc}
\hline
 & Tot $M_{\rm enc}$ & Halo $M_{\rm enc}$   & Tot $M_{\rm enc}$ & Halo $M_{\rm enc}$  \\ 
Mock & 4 kpc & 4 kpc & 9 kpc & 9 kpc \\
Type & $10^{10}\ M_\odot$ & $10^{10}\ M_\odot$ & $10^{10}\ M_\odot$ & $10^{10}\ M_\odot$ \\
\hline\hline
ELL & 4.21 & 0.53 & 8.59 & 4.10 \\
SPH & 4.29 & 0.61 & 9.22 & 4.73 \\
NFW & 5.04 & 1.36 & 9.52 & 5.00 \\ 
IS  & 3.91 & 0.36 & 7.95 & 3.05 \\
\hline
\end{tabular}
\end{table}

%%%%%%%%%%%%%%%%%%%%%%%%%%%%%%%%%%%%%%%%%%%%%%%%%%
\subsection{Exponential Disc Profile} \label{sec:ExpProf}

A thin exponential disc is described by two parameters: the scale length, $r_d$, and total mass, $M$ (Table \ref{mock}). Its surface mass density is given by 
\begin{equation} \label{ExpDisc-Sigma}
\Sigma = \Sigma_0 \exp(-r/r_d)\,,
\end{equation}
where the central surface mass density is related to $M$ and $r_d$ via 
\begin{equation} \label{ExpDisc-Sigma0}
\Sigma_0 = \frac{M}{2\pi r_d^2}\,.
\end{equation}
Integrating to obtain the enclosed mass yields
\begin{equation}
M(r) = 2 \pi \Sigma_0 r_d^2 \left[1 - \exp(-r/r_d)(1+r/r_d)\right]\,.
\end{equation}
The circular speed of an exponential disc is given by
\begin{equation}
v_d(R)^2 = \pi G \Sigma_0 \frac{R^2}{r_d}\left [I_0\left(\frac{R}{2 r_d}\right)K_0\left(\frac{R}{2 r_d}\right)-I_1\left(\frac{R}{2 r_d}\right)K_1\left(\frac{R}{2 r_d}\right)\right]
\end{equation}
where $I_0$, $K_0$, $I_1$, and $K_1$ are modified Bessel functions \citep{2008gady.book.....B}. Since at larger radii in spiral galaxies there is less visible matter, this profile declines with increasing radius.

%%%%%%%%%%%%%%%%%%%%%%%%%%%%%%%%%%%%%%%%%%%%%%%%%%
\subsection{Burkert Model} \label{sec:BurkProf}

A spherical Burkert halo \citep{2000ApJ...537L...9S} has a density profile given by
\begin{equation} \label{Burkert-rho}
\rho(r) = \frac{\rho_c r_c^3}{(r+r_c) (r^2+r_c^2)}\,,
\end{equation}
where $r_c$ is the core radius and $\rho_c$ is the density in the core (Table \ref{mock}). The enclosed mass is 
\begin{equation}
M(r) = \pi \rho_c r_c^3 \left[ -2\arctan x + 2 \ln(1 + x) + \ln (1 + x^2) \right] ,
\label{massenc_sphburk}
\end{equation}
where $x = r/r_c$.  The corresponding circular velocity then follows from $v_h(r)^2 = GM(r)/r$ or
\begin{equation}
v_h(r)^2 = \frac{\pi G \rho_c r_c^2}{x} \left[ -2\arctan x + 2 \ln(1 + x) + \ln (1 + x^2) \right] .
\end{equation}

To obtain an elliptical model, we assume an oblate spheroid with axis ratio $0 < c \le 1$. \citet{2008gady.book.....B} provide a formula for the equatorial rotation curve of such a model:
\begin{equation}
  v_h(r)^2 = 4\pi G \sqrt{1-e^2} \int_{0}^{r} \frac{m^2 \rho(m)}{\sqrt{r^2-e^2 m^2}}\, dm\,,
\end{equation}
where the eccentricity is $e = \sqrt{1-c^2}$. Evaluating the integral yields
\begin{eqnarray}
v_h(r)^2 &=& \pi G \rho_c r_c^2 c \left\{ \frac{2}{\psi_+} \left[\arctan\left(x\right) + \arctan\left(\frac{e^2}{x + c\psi_+}\right)\right] 
 \right. \nonumber\\
&& + \frac{1}{\psi_+} \ln\left[(x^2+1) \frac{(x + \psi_+)^2}{(x+c\psi_+)^2 + e^4}\right] \nonumber\\
&& \left. + \frac{2}{\psi_-} \ln\left[(x+1) \frac{x + \psi_-}{x + e^2 + c\psi_-}\right]\right\}\,,
\label{burkert1}
\end{eqnarray}
where again $x = r/r_c$ and $\psi_{\pm} = \sqrt{x^2 \pm e^2}$. Note that $\psi_-$ becomes imaginary when $x < e$, but with a little algebra we can rewrite the expression as
\begin{eqnarray}
v_h(r)^2 &=& \pi G \rho_c r_c^2 c \left\{ - \frac{2}{\psi_+} \left[\arctan(x) + \arctan\left(\frac{e^2}{x + c \psi_+}\right)\right] \right. \nonumber\\
&& + \frac{1}{\psi_+} \ln\left[(x^2+1) \frac{(x + \psi_+)^2}{(x + c\psi_+)^2 + e^4}\right] \nonumber\\
&& \left. + \frac{2}{\xi_-}\left[\arctan\left(\frac{\xi_-}{x}\right) - \arctan\left(\frac{c\xi_-}{x + e^2}\right)\right]\right\} ,
\label{burkert2}
\end{eqnarray}
where $\xi_- = \sqrt{e^2 - x^2}$. Note that these expressions reduce to the spherical case when $c = 1$ and $e = 0$. For our models we use $e = 0.5$. In practice, there are numerical instabilities (e.g., cases of $0/0$) near $x=e$. We therefore make a Taylor series approximation when $0.3 < x < 1.3$  (for $c = 0.866$ or $e = 0.5$).

We decided to construct separate spherical and elliptical models only for the Burkert profile because of its `cored' nature, which is more aligned with observations (as opposed to NFW which is `cuspy' and the Isothermal profile which is infinite in extent). 

%%%%%%%%%%%%%%%%%%%%%%%%%%%%%%%%%%%%%%%%%%%%%%%%%%
\subsection{NFW Profile} \label{NFWProf}

A spherical NFW halo is specified by two parameters, which can be taken to be the virial mass, $M_{\rm vir}$, and the characteristic density, $\rho_s$ (Table \ref{mock}).  The density profile has the form
\begin{equation} \label{NFW-rho}
\rho = \frac{\rho_s}{(r/r_s)(1+r/r_s)^2}\,,
\end{equation}
where $r_s$ is the scale radius and $\rho_s$ is the scale density. Following \citet{2004MNRAS.351..903G}, we can relate the scale radius to the virial mass as
\begin{equation}
r_s \simeq 5.7\left(\frac{M_{\rm vir}}{10^{11}M_{\odot}}\right)^{0.46} {\rm kpc}\,.
\label{rs}
\end{equation}
Equivalently, if we define the concentration to be $c_s = r_{\rm vir}/r_s$, we can then write
\begin{equation}
c_s \simeq 20\left(\frac{M_{\rm vir}}{10^{11}M_{\odot}}\right)^{-0.13}\,.
\end{equation}
The characteristic density of the distribution, $\rho_s$, is related to the critical density of the universe, $\rho_c$ \citep[see][]{2004MNRAS.351..903G}:
\begin{equation}
\rho_s \simeq \frac{101}{3}\left[\frac{c_s^3}{\ln(1+c_s)-c_s/(1+c_s)}\right]\rho_c\,.
\label{rhos}
\end{equation}
The enclosed mass and rotation curve are:
\begin{eqnarray}
M(r) &=& 4 \pi \rho_s r_s^3\left[\ln\left(\frac{r_s + r}{r_s}\right) -\frac{r}{r_s+r}\right]\,, \\
\label{massenc_NFW}
v(r)^2 &=& 4\pi G \rho_s r_s^3\left[\frac{1}{r} \ln\left(1+\frac{r}{r_s}\right)-\frac{1}{r_s + r}\right] .
\end{eqnarray}

%%%%%%%%%%%%%%%%%%%%%%%%%%%%%%%%%%%%%%%%%%%%%%%%%%
\subsection{Isothermal Sphere} \label{sec:ISProf}
The pseudo-isothermal sphere density profile has density
\begin{equation} \label{IS-rho}
\rho(r) = \frac{v_c^2}{4\pi G} \frac{1}{r_c^2 +r^2}\,,
\end{equation}
where $v_c$ is the asymptotic circular velocity, and $r_c$ is the core radius.  Equivalently, the central density is $\rho_0 = v_c^2/(4\pi G r_c^2)$.  The enclosed mass and rotation curve are:
\begin{eqnarray}
M(r) &=& 4 \pi \rho_0 r_c^3 \left(\frac{r}{r_c} - \arctan\frac{r}{r_c}\right)\,,
\label{massenc_is} \\
v(r)^2 &=& 4\pi G \rho_0 r_c^2\left(1-\frac{r_c}{r} \arctan\frac{r}{r_c}\right) .
\end{eqnarray}
For our purposes, the model is specified by the characteristic density, $\rho_0$, and the core radius, $r_c$ (Table \ref{mock}).

%%%%%%%%%%%%%%%%%%%%%%%%%%%%%%%%%%%%%%%%%%%%%%%%%%
\subsection{Mock Data Production} \label{kinan}

We model large spiral galaxies using radius and mass parameters similar to those of observed galaxies described in \cite{2004MNRAS.351..903G}. Their largest circular velocity was at $\sim$170 km/s, but we increase ours to $\sim$250 km/s to create better candidates for strong lensing. We construct rotation curves by using the parameters in Table \ref{mock} and the equations described above. We assume velocity uncertainties of 3~km/s at each mock data point, similar to typical data \citep[e.g.,][etc.]{1985ApJ...295..305V,2008ApJ...684.1143X,2010ApJ...716..198B,2004MNRAS.351..903G,2010MNRAS.401.1540T}. Our analysis does \textit{not} include random noise, so we may find $\chi^2$ values that are (much) less than the number of degrees of freedom but are still reasonable and meaningful.

%%%%%%%%%%%%%%%%%%%%%%%%%%%%%%%%%%%%%%%%%%%%%%%%%%
\section{Lens Modeling} \label{sec:lensing}
%%%%%%%%%%%%%%%%%%%%%%%%%%%%%%%%%%%%%%%%%%%%%%%%%%

The lensing analysis begins with the lens equation connecting a source at angular position $\vec{u}$ on the sky with an image at angular position $\vec{x}$,
\begin{equation}
\vec{u} = \vec{x} - \nabla \phi(\vec{x})\,,
\end{equation}
where the lens potential $\phi$ is determined from the surface mass density, $\Sigma(\vec{x})$, by the two-dimensional Poisson equation
\begin{equation}
\nabla^2 \phi = 2 \frac{\Sigma}{\Sigma_{\rm crit}}\,.
\label{pois}
\end{equation}
The critical surface density for lensing is
\begin{equation}
\label{sigmacrit}
\Sigma_{\rm crit} = \frac{c^2\,D_s}{4\pi G\,D_l\,D_{ls}}\,,
\end{equation}
where $D_l$ and $D_s$ are angular diameter distances to the lens and source, respectively, while $D_{ls}$ is the angular diameter distance from the lens to the source.

\begin{figure}
\begin{center}$
\begin{array}{c}
\includegraphics[width=75mm]
{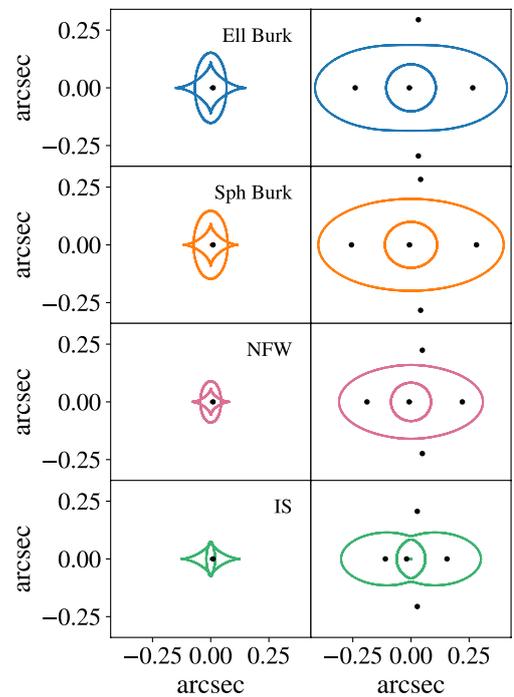}
\end{array}$
\end{center}
\caption{Caustics (left) and critical curves (right) for the mock data fit to the corresponding profile, for the  elliptical Burkert (blue), spherical Burkert (orange), NFW  (purple), and IS model (green). We note that the source position error used was 0.003 arcseconds, the lens position error was 0.005 arcseconds, and the error on the flux was 10\% of the source flux.
\vspace{0.08in}}
\label{causandcritcurves}
\end{figure}

The kinematic parameters from Table \ref{mock} are converted into lensing parameters as described below and listed in Table \ref{mocklens}. Given the  parameters, we use \textit{lensmodel} \citep{2001astro.ph..2340K} to solve the lens equation and compute lensing critical curves and caustics. We place the source near the origin, $\vec{u} = (0.01,0.0)$ arcsec, to obtain a cross image configuration. The images, critical curves, and caustics are shown in Figure \ref{causandcritcurves}. Note that the models predict five images because all of them have a finite central density that leads to a non-vanishing radial critical curve.

In the following subsections we describe the parameter conversions.  Expressions for the lensing potential, deflection, etc.\ can be found in \citet{2001astro.ph..2341K}. 

\begin{table}
\centering
\caption{\label{mocklens}
The lensing parameters were converted from the rotation curve parameters in Table~\ref{mock}. The lens galaxy has redshift $z_l = 0.3$ while the source has redshift $z_s = 2.0$.
}
\begin{tabular}{lcrcl}
\hline
Models & $r_d$ ($''$) &  $\kappa_0$ & $r_c$ ($''$) & $\kappa_h$ \\
\hline\hline
Elliptical Burkert & 0.164  & 1.704 &1.639   &  0.152\\
Spherical Burkert & 0.164  & 1.704 &1.639   &  0.152  \\
\hline
& $r_d$ ($''$)& $\kappa_0$ & $r_s$ ($''$) & $\kappa_s$  \\
\hline
NFW &  0.164  & 1.704 & 1.876  &  0.0743 \\
\hline 
 & $r_d$ ($''$)& $\kappa_0$ & $b_{\rm IS}$ ($''$)  & $s_{\rm IS}$ ($''$) \\
\hline 
  IS & 0.205 & 1.091 & 0.014 & 0.039 \\
\hline
\end{tabular}
\end{table}

%%%%%%%%%%%%%%%%%%%%%%%%%%%%%%%%%%%%%%%%%%%%%%%%%%
\subsection{Exponential Disc Profile}

We imagine viewing the exponential disc from eq.~(\ref{ExpDisc-Sigma}) at an inclination angle $i$ defined such that $i=0$ for a face-on disc while $i=90^\circ$ for an edge-on disc. Then the projected axis ratio (assuming a thin disc) is $q = \cos i$. We adopt $i=60^\circ$ and hence $q=0.5$. For lensing, the disc scale radius $r_d$ is converted to angular units appropriate to the redshift of the lens, while the central surface mass density goes into the dimensionless lensing strength
\begin{equation}
\kappa_0 = \frac{\Sigma_0}{\Sigma_{\rm crit}}\,,
\end{equation}
where $\Sigma_0$ is given by eq.~(\ref{ExpDisc-Sigma0}).

%%%%%%%%%%%%%%%%%%%%%%%%%%%%%%%%%%%%%%%%%%%%%%%%%%
\subsection{Burkert Model}

When the Burkert model from eq.~(\ref{Burkert-rho}) is projected for lensing, the important parameter is the dimensionless lensing strength
\begin{equation}
\kappa_h = \frac{\rho_c r_c}{\Sigma_{\rm crit}}\,.
\end{equation}
For the spherical model, the projection is circular. For the elliptical model, the projection is an ellipse whose projected axis ratio is given by 
\begin{equation}
q = \sqrt{\cos^2 i + c^2\sin^2 i}\,,
\end{equation}
where $i$ is again the inclination angle and $c$ is the semi-axis in the $z$-direction. We fix $i=60^\circ$ as for the disc, and we adopt $c=0.5$ as for the kinematic analysis, which yields a projected axis ratio $q=0.66$.

%%%%%%%%%%%%%%%%%%%%%%%%%%%%%%%%%%%%%%%%%%%%%%%%%%
\subsection{NFW Profile}

The projection of the NFW model from eq.~(\ref{NFW-rho}) is characterized by the dimensionless lensing strength
\begin{equation} \label{ks}
\kappa_s = \frac{\rho_s r_s}{\Sigma_{\rm crit}}\,.
\end{equation}
In our analysis the NFW halo is spherical so its projection is circular.

%%%%%%%%%%%%%%%%%%%%%%%%%%%%%%%%%%%%%%%%%%%%%%%%%%
\subsection{Isothermal Sphere}

The isothermal sphere lens is usually characterized by its Einstein radius and core radius. Given the asymptotic circular velocity $v_c = \sqrt{4\pi G \rho_c r_c^2}$, the Einstein radius parameter is
\begin{equation}
b_{\rm IS} = 2 \pi \left(\frac{v_c}{c}\right)^2 \frac{D_{ls}}{D_{s}}\,.
\end{equation}
The core radius is simply converted to angular units as $s_{\rm IS} = r_c/D_l$. In our analysis the isothermal halo is spherical so its projection is circular.

%%%%%%%%%%%%%%%%%%%%%%%%%%%%%%%%%%%%%%%%%%%%%%%%%%
\subsection{Mock Data Production}

The lensing analysis depends on cosmological distances, which we compute assuming a flat $\Lambda$CDM cosmology with $\Omega_m$ = 0.3, $\Omega_v$ = 0.7, and $H_0=$ 75 km/s/Mpc,   We choose a lens redshift $z_l=0.3$ and source redshift $z_s=2.0$ that are typical of galaxy lensing studies \citep[e.g.,][]{2000ApJ...533..194M,2004IAUS..220..439H,2006MNRAS.370.1008M,2010MNRAS.407..891H}. The critical density for lensing is then 0.525 g cm$^{-2}$. We include typical errors for the image positions (0.003 arcsec) and the galaxy position (0.005 arcsec), and a flux error that corresponds to 10\% of the source flux (see, e.g., Tables 4--5 of \citealt{2018arXiv180709278S}).

%%%%%%%%%%%%%%%%%%%%%%%%%%%%%%%%%%%%%%%%%%%%%%%%%%
\section{Fitting Methods} \label{sec:fitting}
%%%%%%%%%%%%%%%%%%%%%%%%%%%%%%%%%%%%%%%%%%%%%%%%%%

Given a set of mock data, we can fit a model for the mass distribution using standard Bayesian methods. The posterior probability distribution for the model parameters (denoted by $\eta$) given the data (denoted by $d$) can be written as
\begin{equation}
P(\eta|d) = \frac{\overbrace{\textstyle P(d|\eta)}^\text{likelihood} \overbrace{\textstyle P(\eta)}^\text{prior}}{\underbrace{\textstyle P(d)}_\text{evidence}}\,.
\end{equation}
Here the likelihood is $P(d|\eta) \propto e^{-\chi^2/2}$ where the usual goodness of fit is
\begin{equation}
  \chi^2 = \sum_{i=1}^{N} \frac{[]d^{\rm mod}_i(\eta) - d_i]^2}{\sigma_i^2}\,.
\end{equation}
For the kinematic analysis we compute $\chi^2$ directly in python, while for the lensing analysis we compute it using \textit{lensmodel}. We assume flat priors $P(\eta) = \mbox{const}$. We sample the posterior using Markov Chain Monte Carlo methods implemented in the python package \textit{emcee}.\footnote{http://dan.iel.fm/emcee/current/} As noted above, we choose to compare models using enclosed mass. We use the MCMC samples to compute mass values, and we report the median as well as the 68\% confidence interval (spanning the 16th to 84th percentiles). 

We use mock data from each of the four mass distributions, and we fit every data set using all four models. Thus we have a total of sixteen fits each for kinematics and lensing. Having all possible comparisons allows us to examine both statistical and systematic errors in the modeling analysis. We also consider simultaneous fits to the kinematics and lensing data (which can be done simply by summing $\chi^2$ values, since the data are independent).

It is useful to quantify the number of degrees of freedom. Each mass model has four free parameters. Each kinematics data set has $9$ data points, so ${\rm DOF} = 5$. Each lensing data set has $15$ observables (position and brightness for each of five images) along with three additional free parameters (the position and brightness of the source), so ${\rm DOF} = 8$.

%%%%%%%%%%%%%%%%%%%%%%%%%%%%%%%%%%%%%%%%%%%%%%%%%%
\section{Results} \label{sec:results}
%%%%%%%%%%%%%%%%%%%%%%%%%%%%%%%%%%%%%%%%%%%%%%%%%%

\begin{figure}
\begin{center}
\includegraphics[width=85mm]
{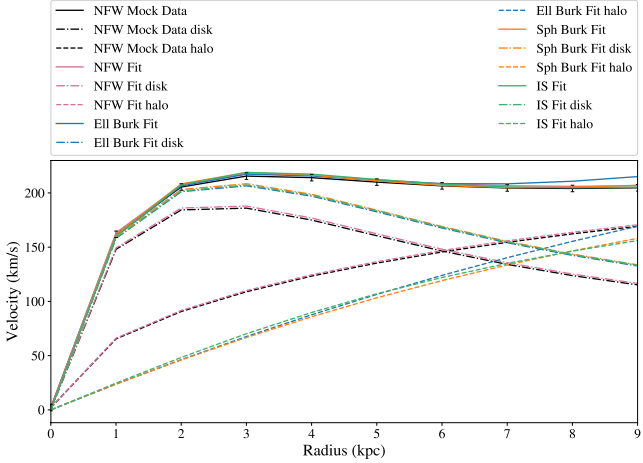} 
\end{center}
\caption{\label{dischalodeg} We illustrate of the common disc/halo degeneracy for spiral galaxy rotation curves. When we fit NFW mock data (black) with various models, it is clear that the isothermal and Burkert models underestimate the halo and overestimate the disc.}
\end{figure}

\begin{figure}
\begin{center}$
\begin{array}{cc}
\includegraphics[width=80mm]{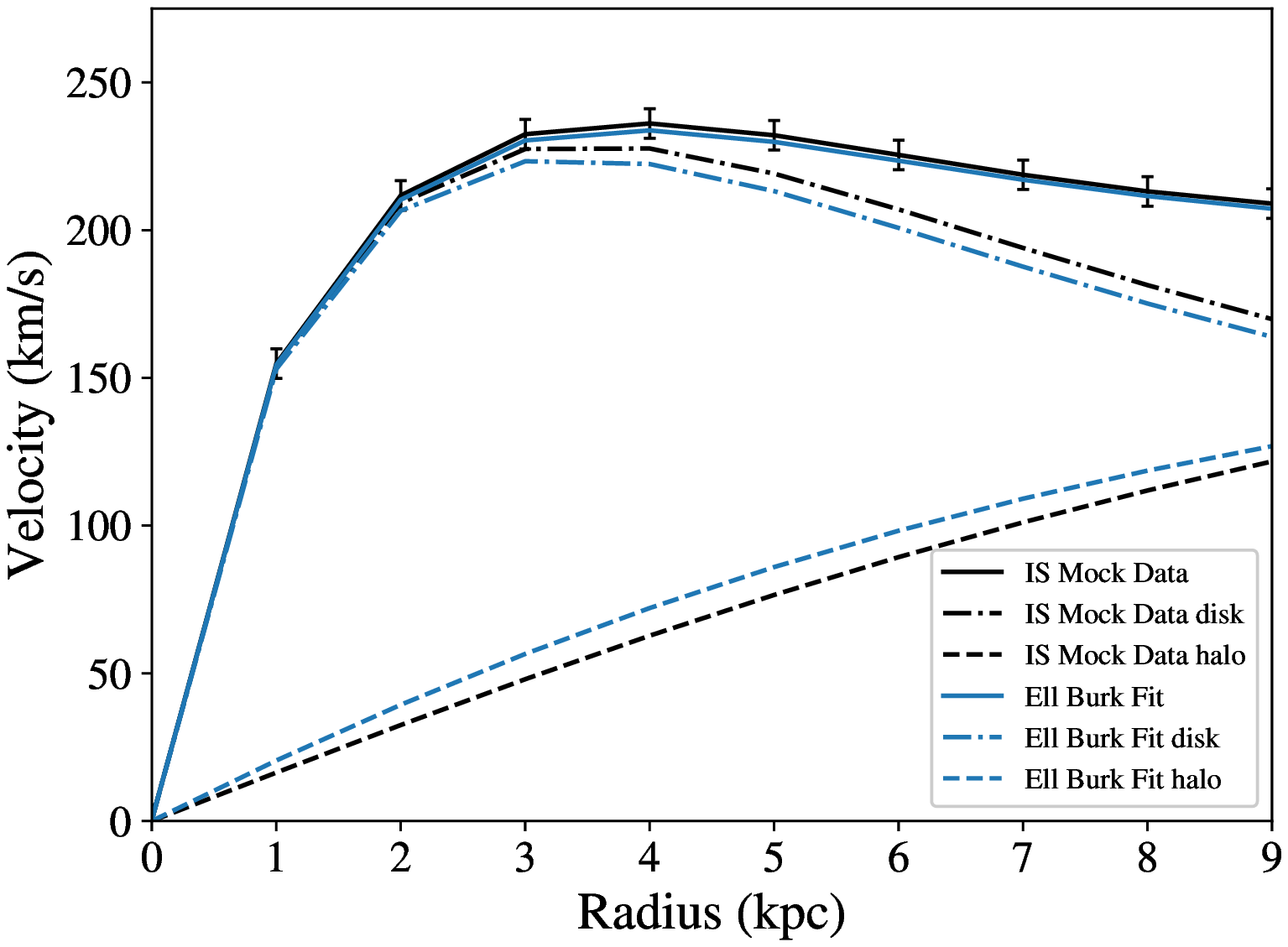} \\
\includegraphics[width=80mm]{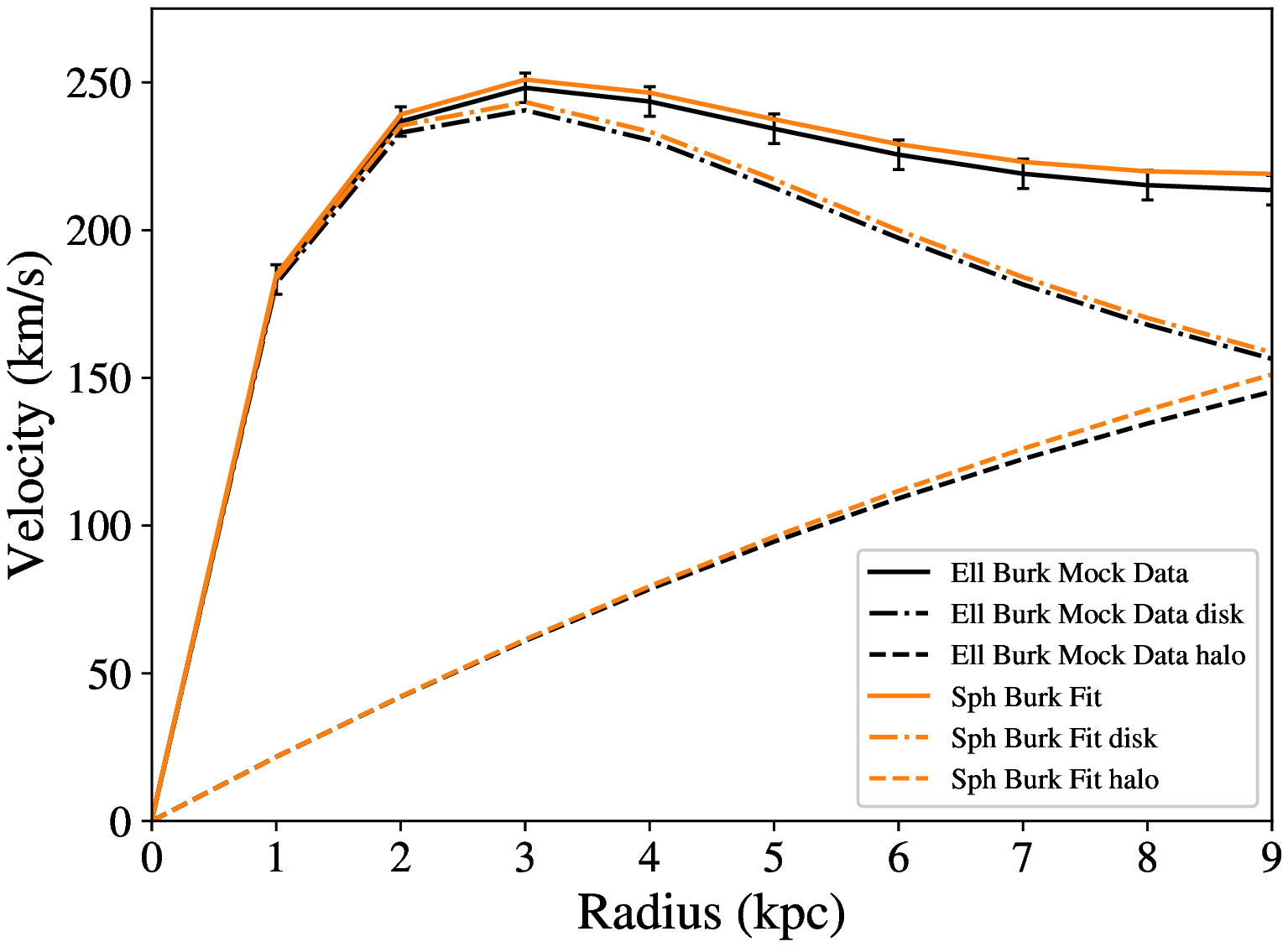}
\end{array}$
\end{center}
\caption{\label{Ex_rotcurve} Examples of rotation curves for two separate scenarios that occur in our kinematic fitting results. \textit{Top:} Here an elliptical Burkert model fit to IS mock data. In this case the fit is statistically good and the original enclosed mass value is correctly reproduced: the true enclosed mass is $7.95 \times 10^{10}\ M_\odot$ and the fit gives an enclosed mass of $7.82\SPSB{+0.21}{-0.20} \times 10^{10}\ M_\odot$, out to 9~kpc. \textit{Bottom}: Here a spherical Burkert model is fit to elliptical Burkert mock data. This is an example of the interesting cases in which we have a statistically good fit yet the enclosed mass values are not consistent: the fit yields $9.20\SPSB{+0.25}{-0.23} \times 10^{10}\ M_\odot$ whereas the true value is $8.59 \times 10^{10}\ M_\odot$. }
\end{figure}

Our fits reveal the common disc/halo degeneracy, as illustrated in Figure\ \ref{dischalodeg}: all four halo models can give reasonable fits to the NFW mock data, but the isothermal and Burkert models underestimate the halo and overestimate the disc. A perhaps more interesting set of results is shown in Figure \ref{Ex_rotcurve}: sometimes models fit the rotation curve and give accurate values for the enclosed mass (top panel), but other times models that are consistent with the rotation curve give incorrect masses (bottom panel). The latter case represents the type of bias we wish to identify.

\begin{figure*}
\centering 
\begin{center}
\includegraphics[width=170mm]{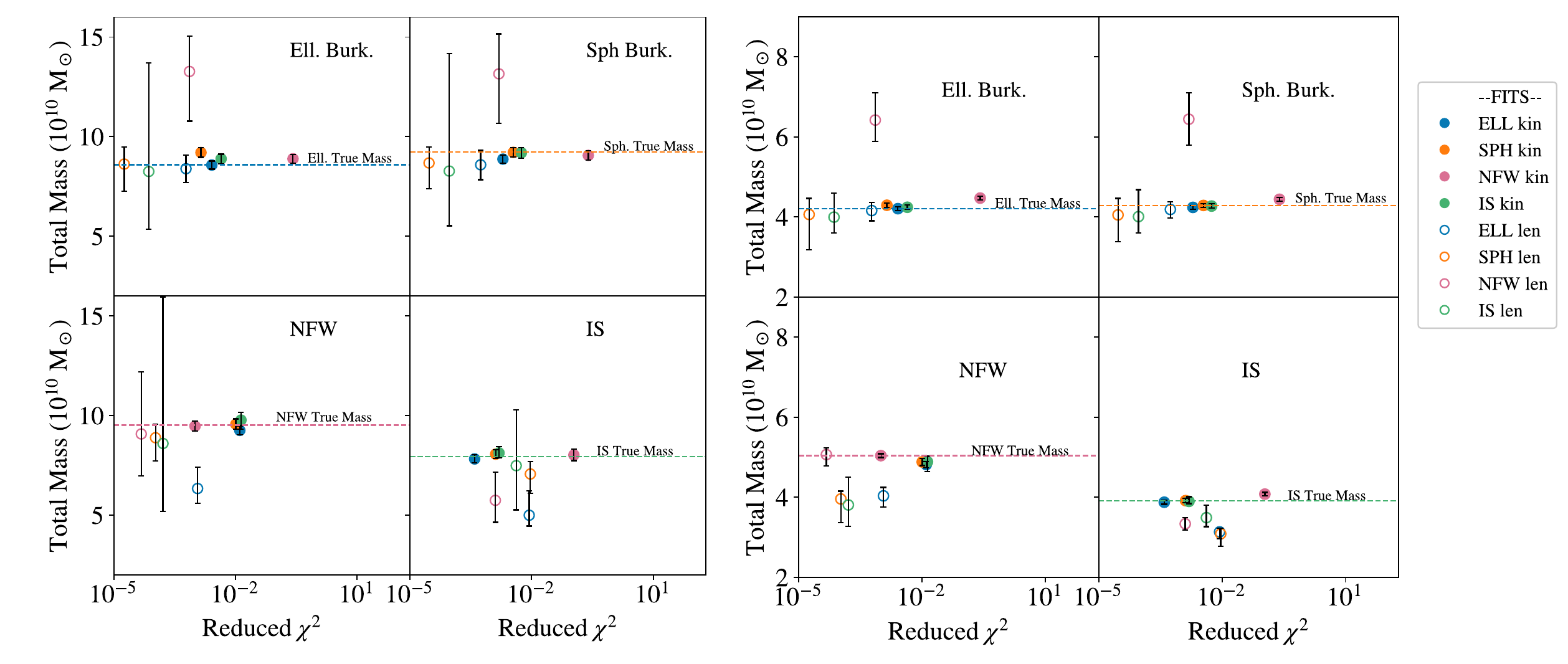} 
\caption{\label{fit_results_1} Visual summary of all the fitting results comparing the {total enclosed mass} out to {9~kpc} ({\em left})  versus the reduced $\chi^2$ of the corresponding fit, and {4~kpc} ({\em right}) versus the reduced $\chi^2$ of that fit. The color indicates the fit to that model, i.e. blue is the elliptical Burkert fit, orange is the spherical Burkert fit, purple is the NFW fit and green is the isothermal sphere fit. The dashed lines represent the true mass for each of the profiles, i.e., fits (points) that lie closer to the dashed line best reproduce the enclosed mass of the true profile. The open circles denote the lensing fits and the closed circles indicate the kinematics fits. In general, the lensing measurements have larger errors and get the enclosed mass wrong more than for the kinematics method.
}
\end{center}
\end{figure*}

\begin{figure*}
\begin{center}$
\begin{array}{c}
\includegraphics[width=170mm]{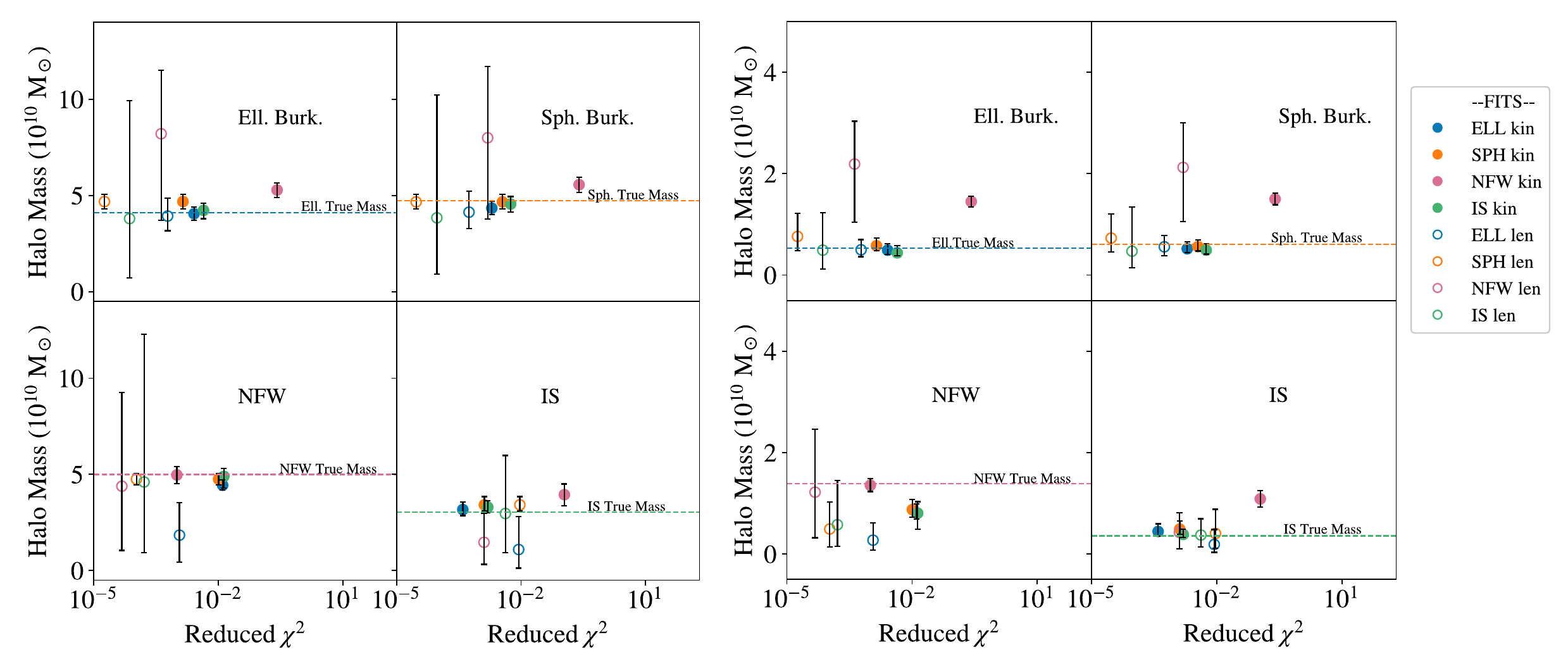} 
\end{array}$
\end{center}
\caption{\label{fit_results_3} Visual summary of the fitting results comparing the enclosed {halo mass} out to {9~kpc} ({\em left})  versus the reduced $\chi^2$ of the corresponding fit, and  {4~kpc} ({\em right}) versus the reduced $\chi^2$ of the fit. Symbols and colors are as in Figure \ref{fit_results_1}. }
\end{figure*}

Table \ref{mockell} reports enclosed mass values from all of the model fits to the elliptical Burkert mock data, for both kinematics and lensing. Tables \ref{mocksph}--\ref{mockIS} then report similar results for the other mock data sets. Values at 4\,kpc and 9\,kpc are chosen as representative probes of the inner and outer regions of the visible galaxy. To better visualise the various results, we plot enclosed mass versus reduced $\chi^2$ for each mock data in Figures \ref{fit_results_1} (for the total enclosed mass) and \ref{fit_results_3} (for the halo mass only). We now discuss the results for kinematics and lensing separately, followed by a joint analysis.
 
%%%%%%%%%%%%%%%%%%%%%%%%%%%%%%%%%%%%%%%%%%%%%%%%%%
\subsection{Kinematics}\label{kinresults}

In the left panel of Figure \ref{fit_results_1} we plot the results of the comparison of the total mass out to 9~kpc (closed symbols).  We find that, for fits to the elliptical Burkert mock data, the spherical Burkert, NFW and IS fits reproduce the rotation curve well but give incorrect enclosed mass values. For example, in Table \ref{mockell} we see that, when the spherical Burkert fit is applied, the enclosed mass out to 9~kpc is $9.20\SPSB{+0.25}{-0.23} \times 10^{10}\ M_\odot$, which is outside the true value of $8.59 \times 10^{10}\ M_\odot$, yet the fitted rotation curve reproduces the data very well (bottom panel of Fig.\ \ref{Ex_rotcurve}). This trend is true as well in the spherical Burkert mock data case and the NFW mock data case for just the elliptical fit. The elliptical Burkert fit to the NFW mock data produces an enclosed mass that, while having small errors, is slightly inconsistent with the enclosed mass produced by the NFW mock data model; more specifically, the elliptical fit to the NFW mock yields an enclosed total mass value of $9.26\SPSB{+0.23}{-0.22} \times 10^{10}\ M_\odot$ out to 9~kpc, whereas the true value is $9.52 \times 10^{10}\ M_\odot$. For the IS mock data model no such case was found. Additionally, we point out that in the NFW mock data case (see Table \ref{mockNFW}), we find that the spherical, elliptical and IS fits underestimate the halo contribution (i.e., the classic disc-halo degeneracy seen clearly in Fig.\ \ref{dischalodeg}). 

In the right panel of Figure \ref{fit_results_3} we plot the results of the comparison of  the total mass out to 4~kpc (closed symbols), and find that in the elliptical, spherical and IS mock data cases the NFW fit reproduces the rotation curve well but yields incorrect enclosed mass values. This is also true in the case of the NFW mock data when fit to either the elliptical and spherical fits. 

We perform the same analysis for the halo mass to see if there are differences or consistencies. In the left panel of Figure \ref{fit_results_3} we plot the results of the comparison of the halo mass out to 9~kpc, and find only one case where the halo rotation curve is reproduced well but the incorrect enclosed mass value is inferred, and this is the spherical fit to the elliptical mock data. In the right panel of Figure \ref{fit_results_3} we plot the outcome of the  comparison of the halo masses out to 4~kpc, and find no discrepancy in the mass values inferred from the fits.

An interesting finding is that, for some configurations, \textit{different models fit the same mock data, yet the enclosed mass values are inconsistent with the actual ones for the models from which the mock data were drawn from.} We have marked with an asterisk in Tables \ref{mockell}, \ref{mocksph}, \ref{mockNFW} and \ref{mockIS} those cases which  provide a good match to the rotation curve, but yield incorrect enclosed mass values. In total there are 11 cases. In Table \ref{combos_kin} we clearly list which combinations yield incorrect mass values but are good fits to the mock data.

%%%%%%%%%%%%%%%%%%%%%%%%%%%%%%%%%%%%%%%%%%%%%%%%%%
\subsection{Lensing}

The results of the fits with the lensing method are also shown in  Figures \ref{fit_results_1} and \ref{fit_results_3} (open symbols). The figures allow to easily visualize the cases for which the lensing method finds the best minimized $\chi^2$ enclosed mass values which are not consistent with the true enclosed mass values, and in Table \ref{combos_lens} we list for which combinations this is true (marked with $^{\dagger}$ in Tables \ref{mockell}, \ref{mocksph}, \ref{mockNFW} and \ref{mockIS}).

\begin{figure}
\begin{center}$
\begin{array}{c}
\includegraphics[width=80mm]{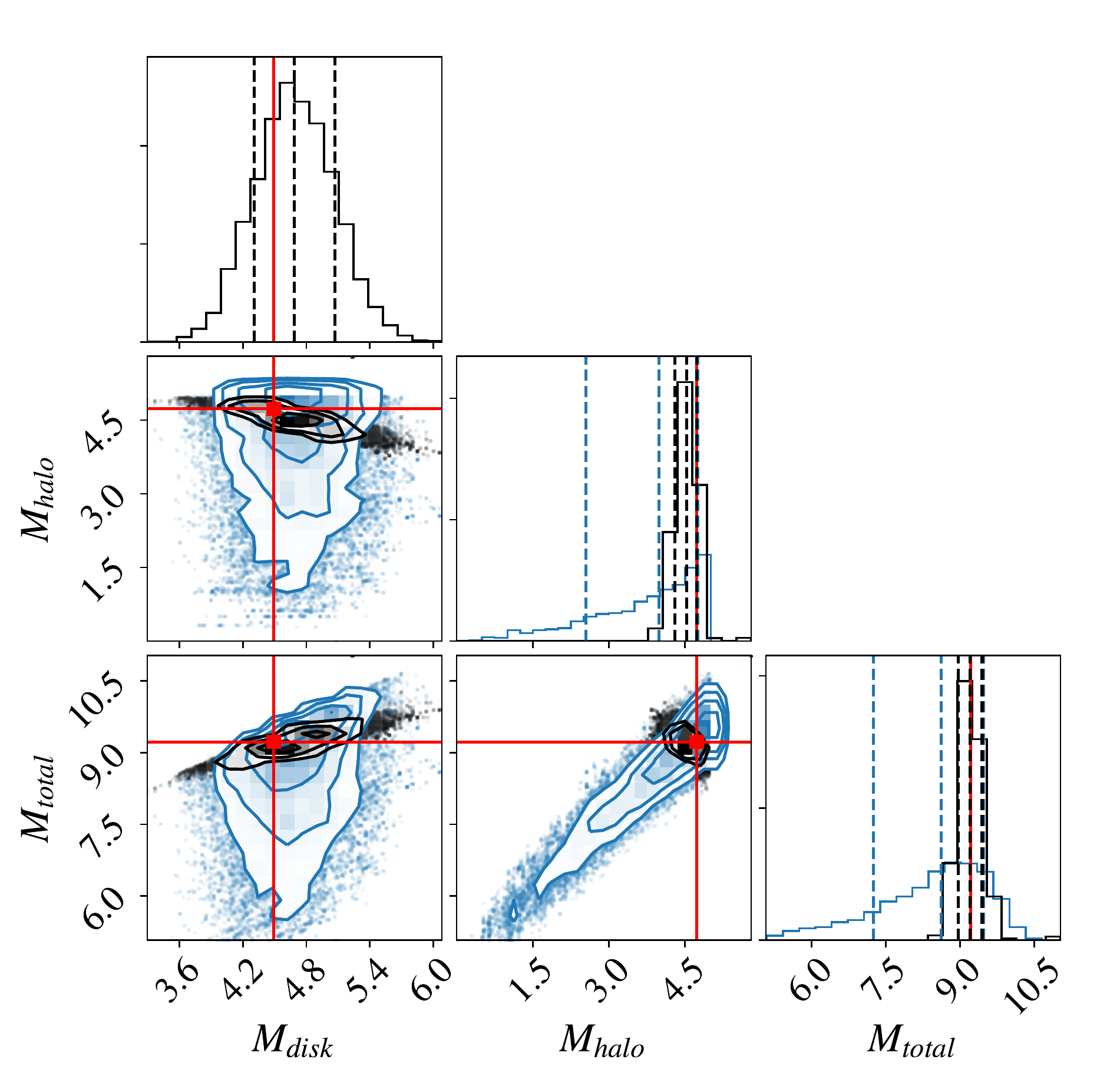}
\end{array}$
\end{center}
\caption{\label{Ellrotcurve_corner}  Posterior probability distribution for the disc, halo, and total masses enclosed within 9~kpc, for kinematics (black) and lensing (blue). Here elliptical Burkert mock data are fit with a spherical Burkert model. The red lines/points indicate the `true' masses (in units of 10$^{10}$ M$_\odot$). It is clear that lensing provides a weaker constraint, which we find to be the case most of the time.} 
\end{figure}

We point out that, for the elliptical Burkert mock data case (Table \ref{mockell}), we find in both the NFW and IS fit that the total mass and the halo mass have large uncertainties at large radii; in particular, the total enclosed mass out to 9~kpc in the NFW fit is $13.27\SPSB{+1.77}{-2.49} \times 10^{10}\ M_\odot$ and in the IS fit is $8.25\SPSB{+5.45}{-2.90} \times 10^{10}\ M_\odot$, yet only the IS fit produces a mass consistent with the true value of $8.59 \times 10^{10}\ M_\odot$. We find that the NFW fit and lensing method together cannot fit the mock data and hence have the largest $\chi^2$ value. In general, we find that kinematics provide better constraints on the parameters (i.e., smaller uncertainties); even though lensing formally has more constraints, they do not appear to be as constraining for the enclosed mass on the scales being considered (see Fig.\ \ref{Ellrotcurve_corner}). Additionally, since we chose baryons to be the dominant constituent in the inner parts of the galaxies, the rotation curve reduces slightly the disc-halo degeneracy. Therefore, this can explain why we find degeneracies less severe than those for lensing. Such mild kinematic degeneracies are not normal for disc galaxies, i.e., most disc galaxies rise smoothly to a plateau (e.g., \citealt{2005ApJ...619..218D}) versus our mock galaxies which drop in amplitude right before the plateau.

We find that even though the $\chi^2$ in the case of the lensing method is smaller, the error bars are significantly larger and the given enclosed mass values are more likely to be inconsistent with the true values. In the IS mock data case (Table \ref{mockIS}) the kinematic method generally does better at constraining each enclosed mass, albeit with larger $\chi^2$ (e.g., for the NFW fit, kinematics gives $\chi^2$ = 0.5 versus lensing which gives $\chi^2$ = 0.01). We remind the reader that we can compare such low $\chi^2$ values because here, by using mock data, we do not sample random noise, unlike in real observational data. 
 
 Last, it is important to keep in mind that the specific $\chi^2$ values that we report depend on our assumed measurement errors. While we adopted typical values, different surveys are made at different sensitivities, and more accurate data will obviously lead towards reducing parameter degeneracies \citep[see, e.g.,][]{2003MNRAS.339..243J}.

%%%%%%%%%%%%%%%%%%%%%%%%%%%%%%%%%%%%%%%%%%%%%%%%%%
\subsection{Combined Analysis}

Next, we couple the kinematics and lensing fits to obtain an overall $\chi^2$. This was done by combining, within the probability function, the likelihood from kinematics with the lensing probability function. 

\begin{figure}%[H]
\begin{center}$
\begin{array}{c}
\includegraphics[width=80mm]{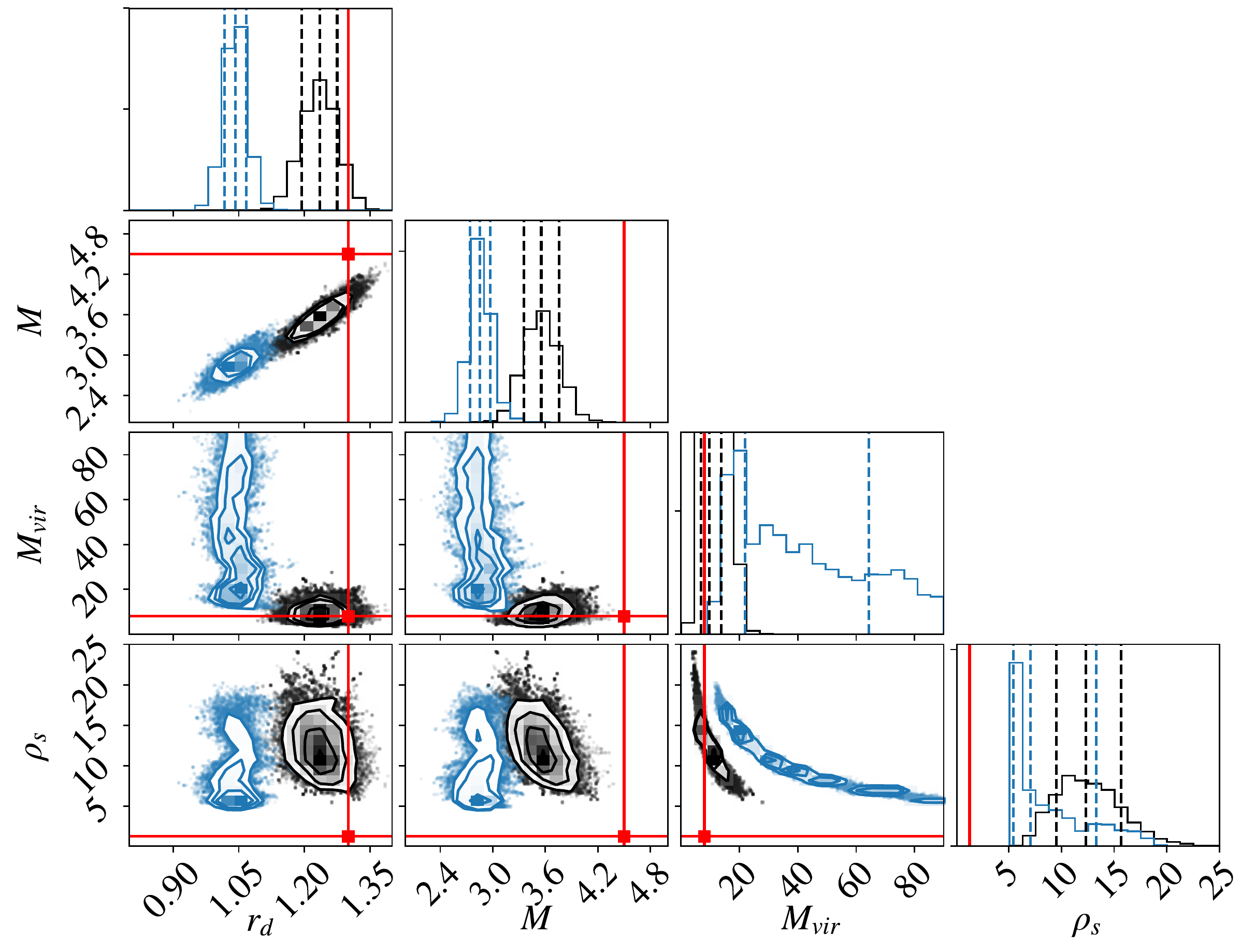}
\end{array}$
\end{center}
\caption{\label{combined_corner} Posterior probability distribution for the model parameters in an NFW fit to spherical Burkert mock data. Here the posterior distributions for kinematics (black) and lensing (blue) do not overlap. The red lines/points indicate the `true' parameters as in Table \ref{mock}.
}
\end{figure}

We often find that the kinematics and lensing data can be fit separately but not jointly. The model that fits the kinematics does not do a good job with the lensing, and vice versa. 
In the case of the NFW fit to the spherical Burkert mock data, for example, we find a large $\chi^2$ of 1440, yet the enclosed mass value is close to the true mass ($9.05 \times 10^{10}\ M_\odot$ vs.\ $9.22 \times 10^{10}\ M_\odot$ ). The large $\chi^2$ value in the joint fit arises because the posteriors from the kinematics and lensing fits have little overlap, as shown in Figure \ref{combined_corner}. Thus a combined analysis of kinematics and lensing may be able to reveal that the model is fundamentally incorrect.

%%%%%%%%%%%%%%%%%%%%%%%%%%%%%%%%%%%%%%%%%%%%%%%%%%
\section{Summary and Conclusions} \label{sec:conclusions}

Dark matter makes up almost a quarter of the energy-density of the Universe; however, its nature remains elusive. Given its dark character,  measurements of the amount of dark matter rely on the  effects of its mass on the surroundings. The two most important methods of measuring mass in individual galaxies are fits to rotation curves and lensing.  In any statistical analysis, it is of paramount importance to be able to assess the robustness with which a statistically good fit with a certain model does indeed provide a realistic description of the data.  Here we have investigated this issue via mock data of rotation curves and lensing, for the four most commonly used halo density profiles (NFW, IS, spherical Burkert and elliptical Burkert).
Since under these conditions the actual model is known by construction, we have been able to assess whether fitting the data with an `incorrect' density profile can sometimes still result in a good fit, and hence yield a biased inference on the amount of DM. 

More specifically, we have been able to provide an estimate of $f_M = M_{\rm fit}/M_{\rm mock}$ for different combinations of `real' profiles data and fitted ones, and learn when and where errors on the inferred $f_M$ are the largest. This analysis has been done with both the kinematic and the lensing methods separately, as well as with combined data, to best mimic the conditions of actual observational surveys.

Table \ref{ratios} summarizes these ratios for the masses out to 9~kpc, for all possible model combinations (four models each fitted with each of the four models), and the two analysis methods, for a total of 32.  
It is evident how, for each set of mock data from a given model, there are some statistically good fits which yield an enclosed mass which can differ by up to about 50\% from the actual values.  The largest discrepancies tend to occur with the lensing method, but there are several cases of inconsistencies also found via the kinematic method.

Some general trends can be identified within our results.  In particular, we note that at large radii the uncertainty in the mass increases, and especially so in those cases in which we are comparing cuspy versus cored profiles.  Interestingly in this regard, an NFW halo was used by \cite{2016MNRAS.463.3151T} in their data modeling; they found good agreement between the lens and dynamical models they used to investigate the mass distribution of a spiral galaxy comparable to ours [$M_{ein} = (7.8\pm 0.3) \times 10^{10}\ M_\odot$]; however, at large radii they noted that the masses inferred with the two methods were becoming inconsistent. In our models, the typical distance of the lensed images from the center of the galaxy is around $\sim$ 1 kpc. Since lensing only constrains the mass within the Einstein radius, this can explain why there is large discrepancy of enclosed mass out to 4 and 9~kpc (columns 6 and 7 in Tables 4-7).

We have compared our mock galaxies to those from the SWELLS survey \citep {2012MNRAS.423.1073B, 2011MNRAS.417.1621D} and found that ours, though smaller on average, show the same trend that the dark matter fraction decreases with circular speed, as shown in Fig.~1 of \citet{2015ApJ...801L..20C}. Also note that the biases we find are within the range of uncertainties of the SWELLS sample.

Overall we find  that, when fitted to mock data from different profiles, the NFW fit underestimates the total mass and overestimates the halo contribution; this holds true for all cases, except of course for the NFW to NFW fit for the kinematic and lensing method, a clear depiction of the disc-halo degeneracy.

To summarize, our work has demonstrated and quantified potential biases in inferring 
the amount of DM mass in spiral galaxies. We have done so independently for the kinematic and the lensing method.
We have found that a tell tale sign of statistically good fits with  incorrect mass measurements could be a discrepancy between the enclosed mass at large radii inferred with the two methods. 

Generally, biases are more pronounced for lensing alone; that is, we find that lensing fits yield more often
incorrect values for the 
 enclosed mass than do kinematics fits alone. Therefore, 
 whenever possible, we encourage observers to add kinematic data in addition to the lensing ones in order to
 correctly retrieve the underlying mass distribution.

%%%%%%%%%%%%%%%%%%%%%%%%%%%%%%%%%%%%%%%%%%%%%%%%%%
\section*{Acknowledgements}
We thank Kristine Spekkens for helpful conversations and comments on the manuscript.
S.~Scibelli's contribution was carried out as part of her senior undergraduate project at Stony Brook Unversity and was in part funded by PSEG as an ``Exploration in STEM" summer student. 

%%%%%%%%%%%%%%%%%%%%%%%%%%%%%%%%%%%%%%%%%%%%%%%%%%



%\bibliographystyle{mnras}


%%%%%%%%%%%%%%%%%%%%%%%%%%%%%%%%%%%%%%%%%%%%%%%%%%

%%%%%%%%%%%%%%%%% APPENDICES %%%%%%%%%%%%%%%%%%%%%
\clearpage 
\newpage

\begin{table*}
\centering
\caption{\label{mockell}
Elliptical Burkert mock data fitting results.
*Kinematic values where the rotation curve fit is statistically good but gives an inconsistent mass value outside given uncertainties. 
$^{\dagger}$Lensing values that give good fit to the mock data but yield an enclosed mass value inconsistent with the true mass.
From Table \ref{mock_encl_mass}, the true enclosed mass values (in units of $10^{10}\ M_\odot$) are as follows:
Total $M_{\rm enc}(4\ {\rm kpc}) = 4.21$,
Halo $M_{\rm enc}(4\ {\rm kpc}) = 0.53$,
Total $M_{\rm enc}(9\ {\rm kpc}) = 8.59$,
Halo $M_{\rm enc}(9\ {\rm kpc}) = 4.10$.
}
\centering
\begin{tabular}{lllcccccc}
\hline 
Mock Data &  Fit &  Method & Tot $M_{\rm enc}$ 4kpc &  Halo $M_{\rm enc}$ 4kpc  & Tot $M_{\rm enc}$ 9kpc & Halo $M_{\rm enc}$ 9kpc & $\chi^2$ &  DOF  \\ & &  &  $10^{10}\ M_\odot$ & $10^{10}\ M_\odot$ &  $10^{10}\ M_\odot$ &  $10^{10}\ M_\odot$ &   &  \\
\hline
\hline
ELL & ELL & Kinematics & $4.21\SPSB{+0.05}{-0.05}$ & $0.49\SPSB{+0.13}{-0.08}$& $8.57\SPSB{+0.20}{-0.21}$ &$4.05\SPSB{+0.34}{-0.34}$ &0.01 &   5  \\
    &     & Lensing & $4.16\SPSB{+0.21}{-0.26}$  & $0.50\SPSB{+0.20}{-0.14}$ & $8.39\SPSB{+0.67}{-0.69}$ & $3.93\SPSB{+0.93}{-0.76}$  & $< 0.01$  &  8   \\
    & SPH &  Kinematics  & $4.29\SPSB{+0.05}{-0.05}$& $0.58\SPSB{+0.15}{-0.10}$& *$9.20\SPSB{+0.25}{-0.23}$ & *$4.68\SPSB{+0.39}{-0.38}$& 0.01  &5   \\
    &  &  Lensing  &  $4.06\SPSB{+0.40}{-0.89}$ & $^{\dagger}0.76\SPSB{+0.45}{-0.28}$ & $8.62\SPSB{+0.86}{-1.37}$ & $4.68\SPSB{+0.39}{-0.38}$ &  $< 0.01$   &   8 \\
    & NFW & Kinematics  & *$4.47\SPSB{+0.04}{-0.04}$& $1.45\SPSB{+0.10}{-0.10}$ &*$8.89\SPSB{+0.23}{-0.22}$& $5.28\SPSB{+0.38}{-0.38}$ &  1.31 &   5    \\ 
    &    & Lensing  & $^{\dagger}$$6.42\SPSB{+0.68}{-0.54}$  & $^{\dagger}$$2.19\SPSB{+0.85}{-1.14}$& $^{\dagger}$$13.27\SPSB{+1.77}{-2.49}$ & $8.21\SPSB{+3.30}{-4.51}$ &  0.01   &  8  \\
    & IS & Kinematics & $4.24\SPSB{+0.06}{-0.05}$ & $0.44\SPSB{+0.14}{-0.06}$& *$8.89\SPSB{+0.24}{-0.25}$ & $4.22\SPSB{+0.38}{-0.43}$ &  0.02 &    5   \\
    &    & Lensing  & $4.00\SPSB{+0.60}{-0.40}$ & $0.49\SPSB{+0.74}{-0.37}$& $8.25\SPSB{+5.45}{-2.90}$ & $3.80\SPSB{+6.13}{-3.09}$  &  $< 0.01$   &  8\\
\hline
\end{tabular}
\end{table*}

\begin{table*}
\centering
\caption{\label{mocksph}
Similar to Table \ref{mockell} but for spherical Burkert mock data.
From Table \ref{mock_encl_mass}, the true enclosed mass values (in units of $10^{10}\ M_\odot$) are as follows:
Total $M_{\rm enc}(4\ {\rm kpc}) = 4.29$,
Halo $M_{\rm enc}(4\ {\rm kpc}) = 0.61$,
Total $M_{\rm enc}(9\ {\rm kpc}) = 9.22$,
Halo $M_{\rm enc}(9\ {\rm kpc}) = 4.73$.
}
\centering
\begin{tabular}{lllcccccc}
\hline 
Mock Data &  Fit &  Method & Tot $M_{\rm enc}$ 4kpc &  Halo $M_{\rm enc}$ 4kpc  & Tot $M_{\rm enc}$ 9kpc & Halo $M_{\rm enc}$ 9kpc & $\chi^2$ &  DOF  \\ & &  &  $10^{10}\ M_\odot$ & $10^{10}\ M_\odot$ &  $10^{10}\ M_\odot$ &  $10^{10}\ M_\odot$ &   &  \\
\hline
\hline
SPH & SPH & Kinematics & $4.29\SPSB{+0.05}{-0.05}$ & $0.56\SPSB{+0.13}{-0.08}$ &$9.21\SPSB{+0.25}{-0.24}$& $4.67\SPSB{+0.40}{-0.36}$ & 0.02  & 5 \\
    &     & Lensing & $4.05\SPSB{+0.41}{-0.67}$ & $0.73\SPSB{+0.47}{-0.27}$& $8.68\SPSB{+0.79}{-1.29}$ & $4.67\SPSB{+0.40}{-0.36}$& $< 0.01$ & 8\\
    & ELL &  Kinematics  & $4.24\SPSB{+0.05}{-0.05}$ & $0.52\SPSB{+0.14}{-0.08}$& *$8.87\SPSB{+0.21}{-0.21}$ & $4.34\SPSB{+0.35}{-0.34}$ &  0.01  &5 \\
    &  &  Lensing  & $4.19\SPSB{+0.19}{-0.22}$& $0.56\SPSB{+0.23}{-0.17}$ & $8.59\SPSB{+0.72}{-0.75}$ & $4.13\SPSB{+1.10}{-0.86}$ & $< 0.01$  & 8\\
    & NFW & Kinematics  &  *$4.44\SPSB{+0.04}{-0.04}$ & $1.50\SPSB{+0.12}{-0.11}$& $9.06\SPSB{+0.23}{-0.23}$ &$5.56\SPSB{+0.40}{-0.40}$&   1.25 & 5 \\ 
    &    & Lensing  & $^{\dagger}$$6.44\SPSB{+0.66}{-0.65}$ & $^{\dagger}$$2.12\SPSB{+0.88}{-1.07}$& $^{\dagger}$$13.15\SPSB{+2.00}{-2.48}$ & $8.00\SPSB{+3.70}{-4.23}$ &  0.01  & 8 \\
    & IS & Kinematics & $4.27\SPSB{+0.06}{-0.05}$ & $0.49\SPSB{+0.13}{-0.08}$ & $9.19\SPSB{+0.25}{-0.28}$ & $4.56\SPSB{+0.39}{-0.41}$ &  0.03 & 5 \\
    &    & Lensing  & $4.01\SPSB{+0.87}{-0.33}$ & $0.47\SPSB{0.87}{0.33}$&  $8.27\SPSB{+5.89}{-2.75}$& $3.84\SPSB{+6.38}{-2.92}$ & $< 0.01$ & 8\\
\hline
\end{tabular}
\end{table*}

\begin{table*}
\centering
\caption{\label{mockNFW}
Similar to Table \ref{mockell} but for spherical NFW mock data.
From Table \ref{mock_encl_mass}, the true enclosed mass values (in units of $10^{10}\ M_\odot$) are as follows:
Total $M_{\rm enc}(4\ {\rm kpc}) = 5.04$,
Halo $M_{\rm enc}(4\ {\rm kpc}) = 1.36$,
Total $M_{\rm enc}(9\ {\rm kpc}) = 9.52$,
Halo $M_{\rm enc}(9\ {\rm kpc}) = 5.00$.
}
\centering
\begin{tabular}{lllcccccc}
\hline 
Mock Data &  Fit &  Method & Tot $M_{\rm enc}$ 4kpc &  Halo $M_{\rm enc}$ 4kpc  & Tot $M_{\rm enc}$ 9kpc & Halo $M_{\rm enc}$ 9kpc & $\chi^2$ &  DOF  \\ & &  &  $10^{10}\ M_\odot$ & $10^{10}\ M_\odot$ &  $10^{10}\ M_\odot$ &  $10^{10}\ M_\odot$ &   &  \\
\hline
\hline
NFW & NFW & Kinematics & $5.04\SPSB{+0.05}{-0.05}$ & $1.36\SPSB{+0.13}{-0.12}$ & $9.47\SPSB{+0.25}{-0.25}$& $4.97\SPSB{+0.44}{-0.45}$ & 0.01  & 5\\
  &  & Lensing & $5.07\SPSB{+0.17}{-0.29}$ & $1.23\SPSB{+1.24}{-0.89}$ & $9.08\SPSB{+3.11}{-2.10}$ & $4.39\SPSB{+4.87}{-3.34}$ & $< 0.01$ & 8\\
  & ELL &  Kinematics  & *$4.82\SPSB{+0.07}{-0.08}$& $0.81\SPSB{+0.18}{-0.12}$ &  *$9.26\SPSB{+0.23}{-0.22}$  &  $4.44\SPSB{+0.29}{-0.25}$& 0.06 &5\\
  &  &  Lensing  & $^{\dagger}$$4.04\SPSB{+0.21}{-0.28}$   & $^{\dagger}$$0.27\SPSB{+0.34}{-0.19}$ & $6.34\SPSB{+1.07}{-0.74}$ & $^{\dagger}$$1.84\SPSB{+1.69}{-1.10}$& 0.01 &8\\
  & SPH & Kinematics  & *$4.88\SPSB{+0.09}{-0.09}$ & $0.88\SPSB{+0.20}{-0.15}$ & $9.56\SPSB{+0.26}{-0.24}$ &  $4.75\SPSB{+0.29}{-0.29}$ & 0.05  &5 \\ 
  &    & Lensing  & $^{\dagger}$$3.96\SPSB{+0.19}{-0.59}$ & $^{\dagger}$$0.49\SPSB{+0.52}{-0.35}$ & $8.89\SPSB{+0.67}{-1.17}$ & $4.75\SPSB{+0.29}{-0.29}$ & $< 0.01$ &8 \\
  & IS & Kinematics &  $4.89\SPSB{+0.14}{-0.24}$ & $0.81\SPSB{+0.23}{-0.32}$& $9.77\SPSB{+0.38}{-0.46}$&  $4.90\SPSB{+0.41}{-0.61}$  & 0.07 &5  \\
  &    & Lensing  & $^{\dagger}$$3.81\SPSB{+0.69}{-0.54}$ & $0.58\SPSB{+0.87}{-0.43}$& $8.60\SPSB{+7.36}{-3.41}$& $4.61\SPSB{+7.68}{-3.68}$  & $< 0.01$ &8\\
\hline
\end{tabular}
\end{table*}

\begin{table*}
\centering
\caption{\label{mockIS}
Similar to Table \ref{mockell} but for spherical IS mock data.
From Table \ref{mock_encl_mass}, the true enclosed mass values (in units of $10^{10}\ M_\odot$) are as follows:
Total $M_{\rm enc}(4\ {\rm kpc}) = 3.91$,
Halo $M_{\rm enc}(4\ {\rm kpc}) = 0.36$,
Total $M_{\rm enc}(9\ {\rm kpc}) = 7.95$,
Halo $M_{\rm enc}(9\ {\rm kpc}) = 3.05$.
}
\centering
\begin{tabular}{lllcccccc}
\hline 
Mock Data &  Fit &  Method & Tot $M_{\rm enc}$ 4kpc &  Halo $M_{\rm enc}$ 4kpc  & Tot $M_{\rm enc}$ 9kpc & Halo $M_{\rm enc}$ 9kpc & $\chi^2$ &  DOF  \\ & &  &  $10^{10}\ M_\odot$ & $10^{10}\ M_\odot$ &  $10^{10}\ M_\odot$ &  $10^{10}\ M_\odot$ &   &  \\
\hline
\hline
IS & IS & Kinematics & $3.90\SPSB{+0.12}{-0.05}$ & $0.39\SPSB{+0.10}{- 0.06}$& $8.12\SPSB{+0.33}{-0.24}$ & $3.30\SPSB{+0.33}{-0.30}$  &  0.01 & 5\\
    &     & Lensing & $^{\dagger}$$3.49\SPSB{+0.31}{-0.22}$ & $0.37\SPSB{+0.32}{-0.23}$ & $7.49\SPSB{+2.81}{-2.22}$ &  $2.97\SPSB{+3.013}{-2.03}$ &  0.03 & 8 \\
    & ELL &  Kinematics & $3.88\SPSB{+0.05}{-0.05}$ & $0.45\SPSB{+0.15}{-0.09}$ & $7.82\SPSB{+0.21}{-0.20}$ & $3.17\SPSB{+0.39}{-0.31}$ & $< 0.01$  & 5\\
    &  &  Lensing  &  $^{\dagger}$$3.14\SPSB{+0.09}{-0.18}$ & $0.19\SPSB{+0.29}{-0.16}$  &  $5.00 \SPSB{+1.23}{-0.54}$  &  $^{\dagger}$$1.09\SPSB{+1.72}{-0.97}$ &0.07 & 8 \\
    & SPH & Kinematics  & $3.92\SPSB{+0.06}{-0.05}$& $0.49\SPSB{+0.16}{-0.11}$& $8.06\SPSB{+0.23}{-0.23}$ &  $3.42\SPSB{+0.42}{-0.33}$& 0.01 & 5\\ 
    &    & Lensing  & $^{\dagger}$$3.09\SPSB{+0.14}{-0.31}$& $0.41\SPSB{+0.48}{-0.29}$ & $7.07\SPSB{+0.63}{-0.97}$ & $^{\dagger}$$3.42\SPSB{+0.42}{-0.33}$ & 0.07& 8 \\
    & NFW & Kinematics & *$4.08\SPSB{+0.04}{-0.04}$ & $1.09\SPSB{+0.16}{-0.16}$ & $8.02\SPSB{+0.29}{-0.29}$ & $3.95\SPSB{+0.56}{-0.57}$ & 0.55 & 5 \\
    &    & Lensing  & $^{\dagger}$$3.34\SPSB{+0.15}{-0.15}$ & $0.43\SPSB{+0.39}{-0.32}$ & $5.74\SPSB{1.42}{1.10}$ &  $^{\dagger}$$1.47\SPSB{1.50}{1.15}$  & 0.01 & 8\\
\hline
\end{tabular}
\end{table*}

\clearpage 
\newpage 

\begin{table}
\centering
\caption{\label{combos_kin}
Outliers in the kinematics analysis. These are combinations that give a good fit to the mock data but yield an enclosed mass value inconsistent with the real one.
}
\centering
\begin{tabular}{lccc}
\hline 
Mock Data & Fit & Mass Type & Distance (kpc) \\
\hline
\hline
ELL & SPH & total & 9  \\
    & SPH & halo & 9 \\
    & NFW &  total & 9 \\
    & NFW & total  & 4  \\
    & IS & total  &   9 \\ 
\hline
SPH & ELL & total &  9 \\
    & NFW  & total & 4  \\
\hline
NFW & ELL & total &  9 \\
    & ELL &  total & 4 \\
    & SPH & total & 4 \\ 
\hline
IS & NFW & total & 4 \\
\hline
\end{tabular}
\end{table}

\begin{table}
\centering
\caption{\label{combos_lens}
Outliers in the lensing analysis.
}
\centering
\begin{tabular}{lllcccc}
\hline 
Mock Data & Fit & Mass Type & Distance (kpc) \\
\hline
\hline
ELL & NFW & total & 9  \\
    & NFW  & total  & 4  \\
    & SPH & halo  &  9 \\ 
        & NFW & halo  &  4\\ 
\hline
SPH & NFW & total &  9 \\
&  NFW & total & 4  \\
&  NFW & halo& 4  \\
\hline
NFW &  ELL & total &  4 \\
& SPH &  total & 4 \\
& IS & total & 4 \\ 
& ELL & halo & 9 \\ 
& ELL & halo & 4 \\ 
& SPH & halo & 4 \\ 
\hline
IS &  NFW & total & 4 \\
&  ELL & total & 4 \\
&  SPH & total & 4 \\
&  IS & total & 4 \\
&  SPH & halo & 9 \\
&  ELL & halo & 9 \\
&  NFW & halo & 9\\
\hline
\end{tabular}
\end{table}

\begin{table}
\centering
\caption{\label{ratios} 
Fit/mock comparison of the enclosed mass up to 9~kpc, for all the models considered here, and for both methods of analysis.
}
\centering
\begin{tabular}{lllcccc}
\hline 
Mock & Fit & Method & $f_M = M_{\rm fit}/M_{\rm mock}$ \\
\hline
\hline
ELL & ELL & Kinematics & $1.00\SPSB{+0.01}{-0.01}$  \\
    &     & Lensing & $0.98\SPSB{+0.06}{-0.05}$  \\
    & SPH &  Kinematics & $1.02\SPSB{+0.01}{-0.01}$ \\
    &  &  Lensing  & $0.96\SPSB{+0.10}{-0.21}$ \\
    & NFW & Kinematics  &   $1.06\SPSB{+0.01}{-0.01}$ \\ 
    &    & Lensing  & $1.52\SPSB{+0.16}{-0.13}$ \\
    & IS & Kinematics &  $1.01\SPSB{+0.01}{-0.01}$ \\
    &    & Lensing  &  $0.95\SPSB{+0.14}{-0.10}$\\
\hline
SPH & SPH & Kinematics &  $1.00\SPSB{+0.01}{-0.01}$ \\
    &     & Lensing & $0.94\SPSB{+0.10}{-0.16}$  \\
    & ELL &  Kinematics & $0.99\SPSB{+0.01}{-0.01}$  \\
    &  &  Lensing  & $0.98\SPSB{+0.04}{-0.05}$ \\
    & NFW & Kinematics  & $1.03\SPSB{+0.01}{-0.01}$ \\ 
    &    & Lensing  & $1.50\SPSB{+0.15}{-0.15}$ \\
    & IS & Kinematics & $1.00\SPSB{+0.01}{-0.01}$ \\
    &    & Lensing  & $0.93\SPSB{+0.20}{-0.08}$ \\
\hline
NFW & NFW & Kinematics &  $1.00\SPSB{+0.01}{-0.01}$ \\
    &     & Lensing &  $1.01\SPSB{+0.03}{-0.06}$ \\
    & ELL &  Kinematics & $0.96\SPSB{+0.01}{-0.02}$ \\
    &  &  Lensing  & $0.80\SPSB{+0.04}{-0.06}$ \\
    & SPH & Kinematics  & $0.97\SPSB{+0.02}{-0.02}$ \\ 
    &    & Lensing  & $0.79\SPSB{+0.04}{-0.12}$\\
    & IS & Kinematics & $0.97\SPSB{+0.03}{-0.05}$ \\
    &    & Lensing  &$0.76\SPSB{+0.14}{-0.11}$ \\
\hline
IS & IS & Kinematics & $1.00\SPSB{+0.03}{-0.01}$ \\
    &     & Lensing &  $0.89\SPSB{+0.08}{-0.06}$  \\
    & ELL &  Kinematics & $0.99\SPSB{+0.01}{-0.01}$ \\
    &  &  Lensing  & $0.80\SPSB{+0.02}{-0.05}$   \\
    & SPH & Kinematics  & $1.00\SPSB{+0.02}{-0.01}$  \\ 
    &    & Lensing  & $0.79\SPSB{+0.04}{-0.08}$ \\
    & NFW & Kinematics & $1.04\SPSB{+0.01}{-0.01}$\\
    &    & Lensing  & $0.85\SPSB{+0.04}{-0.04}$\\
\hline
\end{tabular}
\end{table}

%%%%%%%%%%%%%%%%%%%%%%%%%%%%%%%%%%%%%%%%%%%%%%%%%%

% Don't change these lines
\bsp	% typesetting comment
\label{lastpage}
\end{document}